\documentclass[journal,11pt,onecolumn,draftclsnofoot,]{IEEEtran}
\usepackage{cite}
\usepackage{xcolor}
\usepackage[pdftex]{graphicx}
\usepackage{amssymb}
\usepackage{enumitem}
\usepackage{amsmath}
\graphicspath{{./Figures/}}
\DeclareGraphicsExtensions{.pdf,.jpeg,.png}

\ifCLASSINFOpdf
\else
\fi
\usepackage{amsmath}
\usepackage{algorithmic}
\usepackage{array}
\usepackage{bm}

\newcommand{\myfrac}[2]{\displaystyle \frac{#1}{#2}}

\ifCLASSOPTIONcompsoc
  \usepackage[caption=false,font=normalsize,labelfont=sf,textfont=sf]{subfig}
\else
  \usepackage[caption=false,font=footnotesize]{subfig}
\fi
\usepackage{fixltx2e}
\usepackage{stfloats}
\def\etal{{\it et al.~}}

\ifCLASSOPTIONcaptionsoff
  \usepackage[nomarkers]{endfloat}
 \let\MYoriglatexcaption\caption
 \renewcommand{\caption}[2][\relax]{\MYoriglatexcaption[#2]{#2}}
\fi
\usepackage{url}
\begin{document}
\title{A physics-informed neural network for quantifying 
the microstructure  properties of polycrystalline  Nickel
using ultrasound data}
\author{Khemraj~Shukla,~\IEEEmembership{}
        Ameya D. Jagtap,~\IEEEmembership{}
        James L. Blackshire, ~\IEEEmembership{} \linebreak
        Daniel Sparkman, ~\IEEEmembership{}
        and~George Em Karniadakis~\IEEEmembership{}
\thanks{K. Shukla and A. D. Jagtap are with the Division of Applied Mathematics, Brown University, Providence,
RI, 02906 USA e-mail: (khemraj\_shukla@brown.edu, ameya\_jagtap@brown.edu).}%

\thanks{J. Blackshire and D. Sparkman are with Air Force Research Laboratory, Wright-Patterson AFB, OH, 45433, USA e-mail: (james.blackshire@us.af.mil, daniel.sparkman.1@us.af.mil).}
\thanks{G. Karniadakis is with the Division of Applied Mathematics and School of Engineering, Brown University, Providence,
RI, 02906 USA and PNNL, Richland, WA, 99354 USA e-mail: (george\_karniadakis@brown.edu).}}
\markboth{}%
{Author1 \MakeLowercase{\textit{et al.}}: PINN and Microstructure}

\maketitle
\begin{abstract}
We employ physics-informed neural networks (PINNs) to quantify the microstructure of a polycrystalline Nickel by computing the spatial variation of compliance coefficients (compressibility, stiffness and rigidity) of the material. 
 The PINN is supervised with realistic ultrasonic surface acoustic wavefield data acquired at an ultrasonic frequency of $5~\text{MHz}$ for the polycrystalline material. The ultrasonic wavefield data is represented as a deformation on the top surface of the material with the deformation measured using the method of laser vibrometry. The ultrasonic data is further complemented with wavefield data generated using a finite element based solver. The neural network is physically-informed by the in-plane and out-of-plane elastic wave equations and its convergence is accelerated using adaptive activation functions. 
The overarching goal of this work is to infer the spatial variation of compliance coefficients of materials using PINNs, which for ultrasound involves the spatially varying speed of the elastic waves. More broadly, the resulting PINN based surrogate model shows a promising approach for solving ill-posed inverse problems, often encountered in the non-destructive evaluation of materials.
\end{abstract}


\section{Introduction}
\IEEEPARstart{I}{n} recent years, the availability of large datasets combined with sophisticated algorithms along with the exponential growth in computational power have led to an unprecedented surge of interest in machine learning techniques. Machine learning has been extensively used across the spectrum of disciplines ranging from the classification problem including speech recognition, natural language processing and computer vision, to complex regression problems like approximation of nonlinear and discontinuous functions. However, the applications of neural networks are less explored in the engineering fields.
Physics-informed machine learning approaches define a new paradigm for bridging physical laws with observational data. Recently, such machine learning based techniques have attracted a lot of attention around the world, see \cite{raisi5} and references therein.
In particular, Raissi \etal \cite{raisi5} proposed the physics-informed neural networks (PINNs) methodology, which can accurately solve the forward problem of inferring the solutions of governing physical laws, as well as inverse problems, where free parameters can be inferred or missing physics can be recovered using noisy, sparse and multi-fidelity scattered data sets. The PINN can easily incorporate all the given information such as governing differential/integro-differential equations, experimental/synthetic data etc. into the loss function, thereby recasting the original problem into an optimization problem. Incorporation of governing laws can drastically constrain the space of all admissible solutions. A major advantage of this method is providing a mesh-free algorithm for computation of the differential operators in the governing PDEs, which are approximated by automatic differentiation \cite{baydin}. In this study, we explore the applicability of PINNs in solid mechanics, focusing on the quantification of microstructure in a material based on local elastic property variations. 

The elastic properties of materials such as rigidity $(c_{44},~c_{12})$, and compressibility $(c_{11})$, computed at the microscale level, are utilized in ultrasound quantifying non-destructive sensing to characterize materials. The coefficients $(c_{11}, c_{12}, \text{and}~c_{44})$ are known as elements of the stiffness tensor $\bm{C}$ or compliance coefficients for linear elastic materials. These coefficients are primarily inferred from acoustic and ultrasonic signal data. In a typical measurement setting, an initiating pulse of ultrasonic energy is set to propagate along the surface of a material, which results in oscillation of particles described by the displacement vector $\bm{u}=[u_1, u_2, u_3]$. The $\bm{u}$ is measured in the spatio-temporal domain as surface displacement of particles along the three orthogonal axes. In general, $u_2$  is measured experimentally by using an appropriate experimental setup, wherein the particle's displacements are recorded in the direction perpendicular to the plane-of-symmetry of the material. This approach is referred to as a surface acoustic wave (SAW) sensing method. In the literature, various experimental setups have been used to measure $u_2$ and subsequently used to infer the material properties. A detailed review of some of the methods used to acquire the displacement data was carried out by Xu \etal \cite{xu}. In addition, there are a number of methods that can be used to infer the microstructure of a material from the surface wavefield data $(u_2)$. For example, the direction dependence of ultrasonic phase velocities, which are related to the mechanical properties of a material, have been used to quantify the anisotropy in materials \cite{ledbetter, seiner}. Seiner  \etal \cite{seiner} used this to study the symmetry of structures in polycrystalline Copper by analyzing the variation of anisotropy in different directions. In brief, methods to compute the microstructure properties of a material can be classified into groups (1) direct methods \cite{kube, mendik, noro}, and (2) inverse methods \cite{sathish, xu}. Both direct and inverse methods are typically limited in handling materials based on single-crystal principles. However, Xu \etal \cite{xu} computed the elastic stiffness $(c_{11}, c_{12}, c_{44})$ by inverting the angular distribution of surface wave velocity by acoustic spectro-microscopy for each grain individually using an iterative approach. 

Existing methods to infer the stiffness tensor are primarily based on analyzing the wavefield data represented by $u_2$. However, the methods to analyze the full wavefield data defined by vector $\bm{u}$ have not been fully explored. Therefore, none of the existing methods address an holistic approach of dealing with the entire wavefield data set to compute the stiffness tensor or microstructure properties. Additionally, the wavefield data acquired experimentally or numerically are often sparse and high-dimensional in nature, and therefore, careful attention is required to express these data while inferring the material properties. 

The first successful implementation of PINNs in non-destructive evaluation was carried out by Shukla \etal \cite{shukla}, where they solved the problem of detecting and characterizing a surface crack in a metal plate of Aluminum alloy. The present study is motivated by the work of Shukla \etal \cite{shukla} but the primary objective here is to quantify the microstructure  properties of polycrystalline Nickel from ultrasound non-destructive evaluation measurement data. The primary significance of this work is in the performance of turbine engine materials, where microstructure states play an important role. In this work, we employed PINNs to quantify the microstructure properties by computing $\bm{C}$. The governing equations are defined by two decoupled system of partial differential equations (PDEs). Each set of PDEs expresses in-plane $(u_1, u_3)$ and out-of-plane particle displacement $(u_2)$ explicitly. 
The grain distribution in the polycrystalline Nickel plate attributes to different material properties defined by the local stiffness tensor (compliance coefficients) $\bm{C}$. Wavefield data over the surface of the metal plate is acquired for different ultrasonic frequencies, spatial resolutions, grains statistics, and grain orientation states by using the laser vibrometry sensing method\cite{bshire}. The experimental data only generates out-of-plane displacement $u_2$, therefore the experimental data is further augmented with the numerical data generated using a finite element based solver and synthetic microstructure states generated using the DREAM3D software package \cite{dream3d}. In this paper, we show that PINNs can compute $c_{44}$ from the experimental data and $c_{11}$, $c_{12}$ from the numerical data. Furthermore, we also show that by using only 10-20 \% of the total acquired data along with usage of the adaptive activation functions \cite{ameya}, the objective of computing the stiffness tensor $\bm{C}$ can be successfully achieved.
The conventional solvers often face difficulties to solve ill-posed inverse problems, but PINNs can be easily employed to solve such problems accurately and efficiently. Our approach is novel as we formulate mathematical models for in- and out-of plane particle displacements, and implement such models in PINNs to solve the inverse problem where the stiffness tensor of polycrystalline material is inferred using only the displacement data obtained from the experiment as well as corresponding auxiliary FEM simulations.

This paper is organized as follows. Section II deals with the definition of the problem and the physics of in-plane and out-of-plane particle motion with appropriate and necessary mathematical concepts. In Section III, we describe the PINN methodology in detail and then explain its application for the quantification of microstructure states. Section IV provides the details of pre-processing of the experimental data, followed by section V where all the results are discussed in detail. Finally, we summarize our findings in section VI.

\begin{figure}[!t]
\centering
\subfloat[]{\includegraphics[trim=0cm 0cm 0cm 1cm, width=0.5\textwidth]{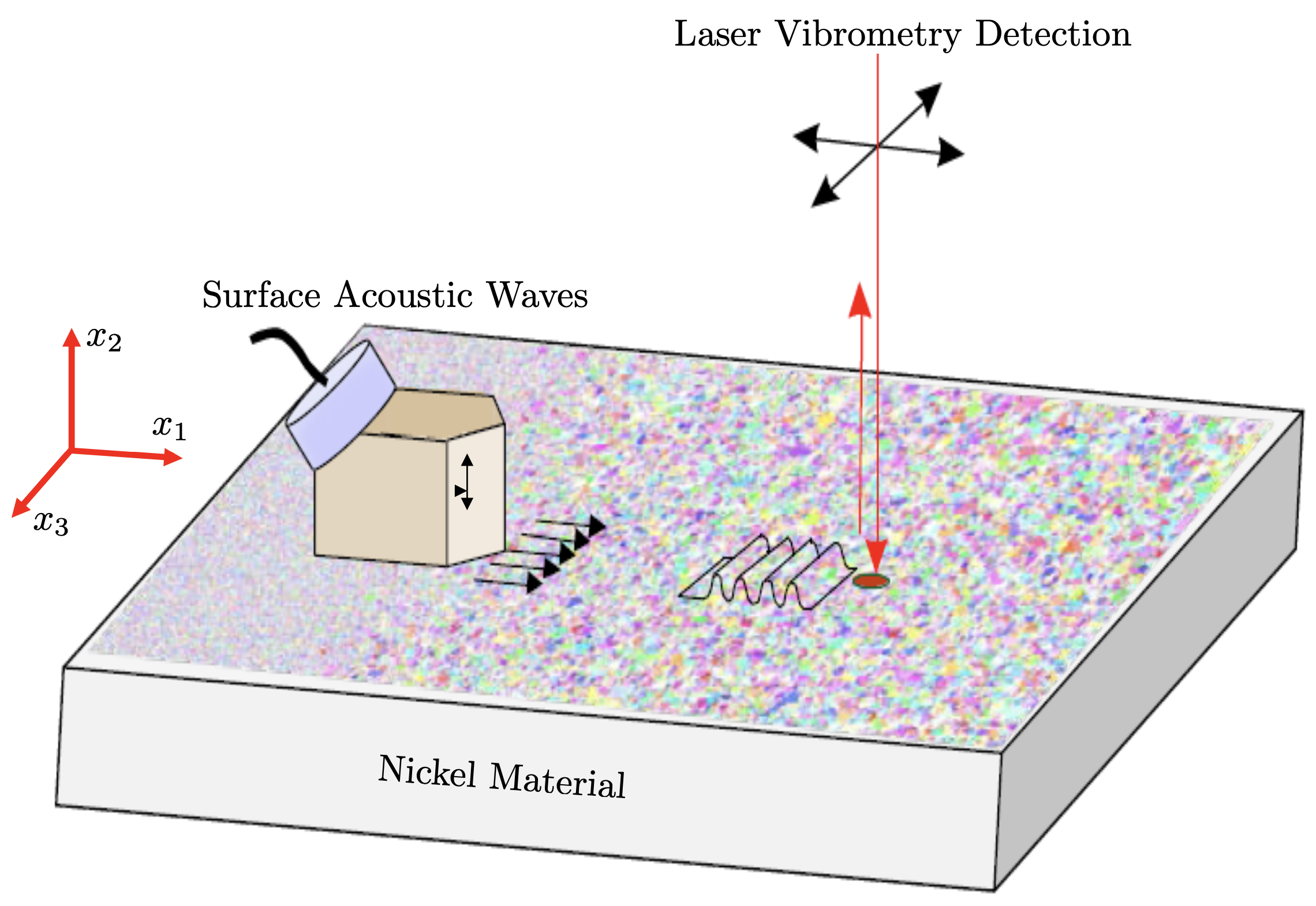}
\label{Figure_wa}}
\subfloat[]{\includegraphics[trim=0cm 0cm 0cm 1cm, width=0.45\textwidth]{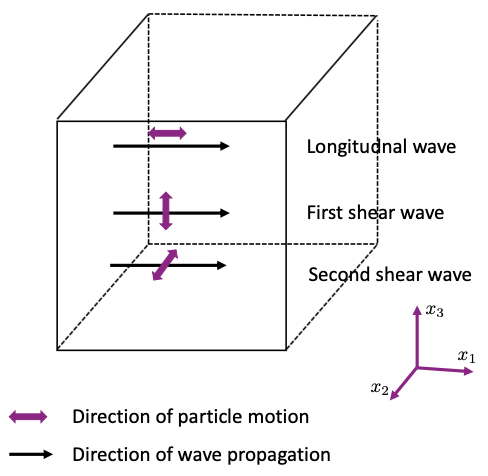}
\label{Figure_wb}}
\caption{(a) Experimental setup for measuring $u_2$ on a metal surface. The color patches on surface represent polycrystalline grain orientation variations. (b) Types of waves and particle motions generated due to propagation of ultrasonic waves.}
\label{fig:fig1}
\end{figure}
\section{Problem Definition}
In this section, we will describe the physics of elastic wave propagation for in-plane and out-of-plane particle motion along with the details on the corresponding data sets. Prior to that, we will briefly describe the definition of the constitutive relations and the specific definitions for polycrystalline Nickel. Polycrystalline Nickel exhibits a cubic symmetry and its microstructure is primarily defined by the elements of stiffness tensor, i.e., $c_{11},~c_{12}$ and $c_{44}$. Here, $c_{11}$ represents the linear combination of compressibility and rigidity, whereas $c_{12}$ and $c_{44}$ define the rigidity of the materials. These coefficients define the strain in a continuum represented by a control volume whose surfaces are parallel to the coordinate system represented by $(x_1,x_2, x_3)$. The relation between stress and strain is expressed as \cite{JMC}
\begin{align} \label{Ctensor}
\bm{\sigma}=\left[\begin{array}{cccccc}
c_{11} & c_{12} &c_{12} & 0 & 0 & 0\\
c_{12} & c_{11} & c_{12} & 0 & 0 & 0\\
c_{12} & c_{12} & c_{11} & 0 & 0 & 0\\
0 & 0 & 0 & c_{44} & 0 & 0\\
0 & 0 & 0 & 0 & c_{44} & 0\\
0 & 0 & 0 & 0 & 0 & c_{44}
\end{array}\right]\bm{\epsilon},
\end{align}
where $\bm{\sigma}=[\sigma_{11}, \sigma_{22},\sigma_{33}, \sigma_{23},\sigma_{13},\sigma_{12}]^T$ and $\bm{\epsilon}=[\epsilon_{11}, \epsilon_{22},\epsilon_{33}, 2\epsilon_{23},2\epsilon_{13},2\epsilon_{12}]^T$ are the stress and strain tensors, respectively. 

The subscripts of notations in (\ref{Ctensor}) follow the Voigt notation \cite{JMC}. For example, in $\sigma_{11}$  the first index specifies the direction in which the stress component acts, i.e., $x_1$, while the second index identifies the orientation of the surface upon which it is acting (which is also $x_1$). The same definition applies to $c_{11},~c_{12}$ and $c_{44}$ and the components of the tensors. Elements of strain tensor ($\bm{\epsilon}$) are expressed in terms of the particle displacements $\bm{u}=[u_1, u_2, u_3]$, which are the measured and simulated data to be used for the quantification of the microstructure. The following is the expression for strain component applicable in the context of this study
\begin{align} \label{strain}
\epsilon_{11} = \myfrac{\partial u_1}{\partial x_1}, \qquad \epsilon_{33} = \myfrac{\partial u_3}{\partial x_3}, \qquad \text{and} \qquad \epsilon_{13} = \myfrac{1}{2}\left(\myfrac{\partial u_1}{\partial x_3} + \myfrac{\partial u_3}{\partial x_1} \right). \end{align}

In Figure 1a, we describe the experimental setup for measuring the particle displacement at the surface of the material by using the propagation of ultrasonic surface acoustic waves. An impulse excitation centered about 5 MHz was induced on the surface of the material. As the wave traverses, energy in the wave causes particles to oscillate. The direction of particle motion vis-a-vis energy propagation is affected by the elastic properties of the material. Figure \ref{fig:fig1} (b) shows the three cases of particle displacement and energy propagation. The first case corresponds to the compressive-extension motions where the particle motion is along the direction of energy propagation and represents longitudinal waves. The second case defines the particle motion being orthogonal to the direction of energy propagation, where particles polarized in the vertical plane represent the first shear wave. Finally, the last case corresponds to particle motions orthogonal to the direction of energy propagation with particles polarized in the horizontal plane representing the second shear wave. 
We denote the velocity of longitudinal wave, first shear wave, and second shear wave as $v_{p}$, $v_{s_1}$ and $v_{s_2}$, respectively. The particle displacement corresponding to longitudinal, first, and second shear waves are denoted as $u_1,~u_3,$ and $u_2$, respectively. To measure $u_2$, we have used the setup shown in Figure \ref{fig:fig1} (a) and the wavefield imaging method \cite{bshire}. In practice, the wavefield imaging method uses a scanning laser vibrometry system to detect displacement fields on a material surface with nanometer precision. The laser vibrometer uses a focused laser beam and coherent light interference principles to convert phase interference of the light into particle displacement surface deformation measurement. For the present study, a contact wedge surface acoustic waves transducer was held stationary on the material surface to generate the ultrasound field Figure 1a, while the laser vibrometry beam was scanned in the $(x_1, x_3)$ plane \cite{bshire}. Repetitions of the ultrasonic excitation for different vibrometry detection points $(x_1, x_3)$ enables the collection of time series data (displacement amplitude vs time signals) at discrete $(x_1, x_3)$ spatial locations and captured at regularized grid points. These measurements are stored as a collection of time-evolving snapshots of $u_2(x_1, x_3, t)$ data showing important details of the ultrasound wave propagation and scattering phenomena occurring between the ultrasound field and the local material features (e.g. grain boundaries).


In a 2D plane-of-symmetry $(x_1-x_3)$, particle displacement $u_2$, contained in a plane perpendicular to $(x_1-x_3)$ plane $(\partial x_2 = 0)$, is expressed as  \cite{JMC}
\begin{align}\label{outplane}
\rho(\bm{x})\myfrac{\partial ^2 u_2}{\partial t^2} = \left(\myfrac{\partial \sigma_{12}}{\partial x_1} + \myfrac{\partial \sigma_{23}}{\partial x_3} \right),
\end{align}
where $\sigma_{12} = c_{44} (\bm{x}) \myfrac{\partial u_2}{\partial x_3}$ and $\sigma_{23} = c_{44} (\bm{x}) \myfrac{\partial u_2}{\partial x_1}$, $\rho(\bm{x})$ is density of material, $x_1, x_3 \in \Omega \subset \mathbb{R}^2$ and $t \in [0, T]$.
Substituting the definition of $\sigma_{12}$ and $\sigma_{23}$ in (\ref{outplane}), we recover
\begin{align}\label{outplane-f}
\myfrac{\partial ^2 u_2}{\partial t^2} = 2\widehat{c}_{44} ~\myfrac{\partial^2 u_{2}}{\partial x_1 \partial x_3},
\end{align}
where $\widehat{c}_{44} = \myfrac{c_{44} (\bm{x})}{\rho(\bm{x})}$.

To compute the $c_{11}$ and $c_{12}$, $u_1$ and $u_3$ are generated numerically by using the DREAM-3D package \cite{dream3d}. The particle displacement $u_1$ and $u_3$ in the $(x_1, x_3)$ plane are defined by following set of PDEs \cite{JMC}
\begin{align}
\begin{aligned}
\label{inplane}
\myfrac{\partial^2 u_1}{\partial t^2}&= \myfrac{\partial \sigma_{11}}{\partial x_1} + \myfrac{\partial \sigma_{13}}{\partial x_3},\\
\myfrac{\partial^2 u_3}{\partial t^2}&= \myfrac{\partial \sigma_{13}}{\partial x_1} + \myfrac{\partial \sigma_{33}}{\partial x_3},
\end{aligned}
\end{align}
where
$
\sigma_{11}= \widehat{c}_{11} \myfrac{\partial u_1}{\partial x_1} + \widehat{c}_{12} \myfrac{\partial u_3}{\partial x_3}, ~\sigma_{33} = \widehat{c}_{12} \myfrac{\partial u_1}{\partial x_1} + \widehat{c}_{11} \myfrac{\partial u_3}{\partial x_3}
~\text{and}~
\sigma_{13} = \widehat{c}_{44}\left(\myfrac{\partial u_3}{\partial x_1} + \myfrac{\partial u_1}{\partial x_3} \right)
$ with $\widehat{c}_{ij} = \myfrac{c_{ij} \bm{(x)}}{\rho{\bm{(x)}}}$.

The speed of longitudinal, first shear and second shear waves in the Nickel is expressed as \cite{miglori} 
\begin{align} \label{speed}
\begin{aligned}
v_{p_1} = \sqrt{\myfrac{c_{11}}{\rho}},\qquad v_{s_1} = \sqrt{\myfrac{c_{11} -c_{12}}{\rho}}, \qquad \text{and}~\qquad v_{s_2} = \sqrt{\myfrac{c_{44}}{\rho}}   
\end{aligned}
\end{align}

Thus, in this paper we will be addressing the following problem. 
Given the displacement vector $\bm{u}=[u_1, u_2, u_3]$ for polycrystalline Nickel, we have to compute $c_{11}$, $c_{12}$ and $c_{44}$ (or $v_{p_1}$, $v_{s_1}$ and $v_{s_2}$) from the neural networks informed by the physics defined in equations (\ref{outplane-f}) and (\ref{inplane}).

\section{Physics-informed Neural Network}
\label{PINN}
In this section we will describe the design of the PINN to compute  $\widehat{c}_{11},~\widehat{c}_{12}$ and $\widehat{c}_{44}$. Prior to that, we shall briefly explain the mathematical setup for the fully-connected feed-forward neural network (FFNN).


\subsection{Neural network - Mathematical setup}
We have chosen a FFNN of depth $D$ with an input layer, $D-1$ hidden layers and an output
layer. In the $k^{\text{th}}$ hidden layer, the number of neurons is denoted by $N_k$. Each hidden layer receives an affine transformed output $\bm{x}^{k-1} \in \mathbb{R}^{N_{k-1}}$ from the previous layer. The affine transformation is expressed as
\begin{align} \label{eq1}
\mathcal{L}_k(\bm{x}^{k-1}):=\bm{w}^k \bm{x}^{k-1} + \bm{b}^k.    
\end{align}
In the $k^{th}$ layer of the network, the weights $\bm{w}^k\in \mathbb{R}^{N_{k} \times N_{k-1}}$ and the biases $\bm{b}^k \in \mathbb{R}^{N_{k}}$ are initially chosen from \textit{independent and identically distributed (i.i.d)} samplings. The nonlinear activation function $\Phi(\cdot)$ is applied to  $\mathcal{L}_k(\bm{x}^{k-1})$ component-wise prior to sending it as input to the next layer. The activation function at the output layer is an identity function. Thus, the final neural network representation is expressed as
\begin{align}\label{eq2}
\bm{u}_\Theta(\bm{x})=\left(\mathcal{L}_k \circ \sigma \circ \mathcal{L}_{k-1} \circ ...\circ \sigma \circ \mathcal{L}_1\right)(\bm{x}),    
\end{align}
where $\bm{u}_\Theta(\bm{x})$ is the output of the neural network and the operator $\circ$ is the composition operator, $\Theta=\{\bm{w}^k, \bm{b}^k, a \}_{k=1}^D$ represents all the trainable parameters in the network including the slope of the activation function defined as $a$. 
Adaptive activation functions are shown to accelerate the training of the deep as well as physics-informed neural networks by introducing the trainable parameters in the activation function, which can be optimized along with the other parameters in the neural networks. We employ global adaptive activation function, see \cite{ameya, ameya1, ameya2} for more details.

\begin{figure}[!t]
\centering
\includegraphics[trim=4cm 4cm 11cm 1cm, scale=0.60]{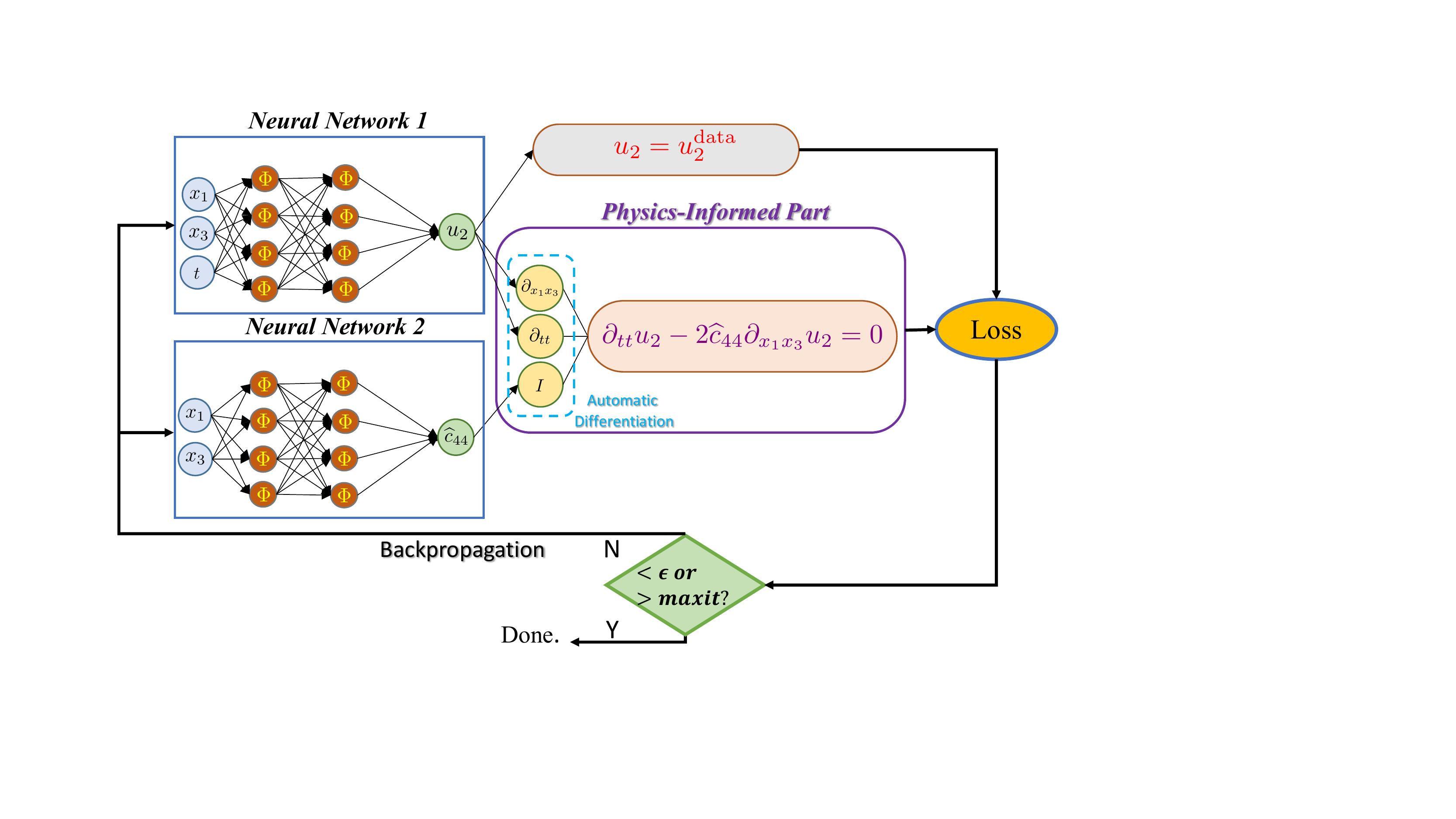}
\caption{Physics-informed neural network to compute $\widehat{c}_{44}$, described  in equation (\ref{outplane}).}
\label{PINN}
\end{figure}

\subsection{PINNs for computing $\widehat{c}_{11}$, $\widehat{c}_{12}$ and $\widehat{c}_{44}$}
In this section, we shall explain the architecture of the neural networks informed by governing equations (equations (\ref{outplane}) and (\ref{inplane})), which will infer $\widehat{c}_{12}$, $\widehat{c}_{12}$ and $\widehat{c}_{44}$. 

\begin{itemize}
    \item \textbf{PINN for} $\mathbf{\widehat{c}_{44}}$:

First, we will define the construction of PINN to compute $\widehat{c}_{44}$ from the particle displacement data $u_2$. We define the residual $f_2$ from (\ref{outplane-f}) as
\begin{align} \label{eq3}
f_2:= \myfrac{\partial^2 u_2}{\partial t^2}- 2\widehat{c}_{44} \left( \myfrac{\partial^2 u_2}{\partial x_1 \partial x_3}\right),
\end{align}
and proceed by approximating  $u_2(t,x)$ and $\widehat{c}_{44}(\bm{x})$ with two separate deep neural networks 
as shown in Figure \ref{PINN} with blocks annotated with \textit{Neural Network 1} and \textit{Neural Network 2}. The reason behind choosing the two separate neural networks to learn $u_2$ and $\widehat{c}_{44}$ is their dependencies on different input variables, which for $u_2$ is $(x_1, x_3, t)$ whereas for $\widehat{c}_{44}$ is $(x_1, x_3)$. It is important to note that during the training procedure both the networks are trained simultaneously.
The output of the neural networks not only satisfies the given data but also the governing equation as shown by the physics-informed part.
The spatio-temporal derivatives in the governing equation (\ref{eq3}) are computed by using the automatic differentiation \cite{baydin}. The PINN algorithm aims to learn a surrogate $u_2 = u_{2_{\Theta}}$ for predicting the solution of (\ref{outplane-f}).
The PINN loss function is given as 
\begin{align} \label{eq5}
J_{\widehat{c}_{44}}(\Theta)= \lambda_{u_2} MSE_{u_2} + \lambda_{f_2} MSE_{f_2},
\end{align}
where the data loss is given by $MSE_{u_2}=\myfrac{1}{N_{u_2}}\sum_{i=1}^{N_{u_2}} \left \vert u_{2_{\Theta}}(\bm{x}_{u_2}) - u_2^{\text{data}}(\bm{x}_{u_2}) \right \vert^2$, and the residual loss is  $MSE_{f_2} = \myfrac{1}{N_{f_2}}\sum_{i=1}^{N_{f_2}} \left \vert {f_2}(\bm{x}_{f_2}) \right \vert^2$. The parameter $\lambda_{u_2},\lambda_{f_2} > 0$ is the penalty parameter, which helps in achieving the fast convergence. Here, $\bm{x}_{u_2}$  and $\bm{x}_{f_2}$ represent the spatio-temporal locations of training data (experimentally measured $u_2$) and the residual points, respectively. Our objective is to achieve the optimal parameters of neural network for which the loss function defined in (\ref{eq5}) is minimized. Thus, the definition of the resulting optimization problem is expressed as
\begin{align} \label{eq6}
\Theta^* = \arg \min_{\Theta \in \mathcal{V} } ~~J_{\widehat{c}_{44}}(\Theta),
\end{align}
where $\mathcal{V}$ is the parametric space.
The stochastic gradient descent (SGD) method is the widely used optimization
method. In SGD, a small set of points is randomly sampled to find the direction of the gradient in every iteration. The SGD algorithm works well to avoid bad local minima during training of DNN under the one point convexity property. In particular, we shall use the Adam optimizer, which is one of the version
of SGD, followed by the L-BFGS optimizer.

\item \textbf{PINNs for} $\mathbf{\widehat{c}_{11}}$ \textbf{and} $\mathbf{\widehat{c}_{12}}$:

For constructing the parameters $\widehat{c}_{11}$ and $\widehat{c}_{12}$,  we shall use the already computed $\widehat{c}_{44}$ from the previously defined PINNs. The residuals for equation \ref{inplane} is expressed as
\begin{align} \label{inplane-resi}
\begin{aligned}
f_1 := \myfrac{\partial ^2 u_1}{\partial t^2}&-\left[\widehat{c}_{11} \myfrac{\partial^2 u_1}{\partial x_1^2} + \widehat{c}_{12} \myfrac{\partial^2 u_3}{\partial x_1 \partial x_3} + \widehat{c}_{44} \left(\myfrac{\partial^2 u_3}{\partial x_3 \partial x_1} + \myfrac{\partial ^2 u_1}{\partial x_3^2} \right)\right]\\ 
f_3 :=\myfrac{\partial ^2 u_3}{\partial t^2}&-\left[\widehat{c}_{12} \myfrac{\partial^2 u_1}{\partial x_1 \partial x_3} + \widehat{c}_{11} \myfrac{\partial^2 u_3}{\partial x_3^2} + \widehat{c}_{44} \left(\myfrac{\partial^2 u_3}{\partial x_1^2} + \myfrac{\partial ^2 u_1}{\partial x_1 x_3} \right)\right]\\ 
\end{aligned}    
\end{align}
The PINN consists of three networks where the first network approximates the data $u_1(\bm{x})$ and $u_3(\bm{x})$. The second and third networks approximate $\widehat{c}_{11}$ and $\widehat{c}_{12}$, respectively. The combination of these three networks approximates $u_1$ and $u_3$, which are approximated at the sampled residual points and fed to the residuals defined in (\ref{inplane-resi}). Therefore, the loss function is defined as
\begin{align} \label{eq8}
J_{\widehat{c}_{11}, \widehat{c}_{12}}(\Theta)= \lambda_{{u_1}} MSE_{u_1} + \lambda_{{u_3}} MSE_{u_3} +  \lambda_{f_1} MSE_{f_1} + \lambda_{f_3} MSE_{f_3},
\end{align}
where $MSE_{u_1}$, $MSE_{u_3}$, $MSE_{f_1}$ and $MSE_{f_3}$ correspond to mean square error of $u_1$, $u_3$, $f_1$ and $f_3$, respectively. Again, the parameters $\lambda_{u_1}$, $\lambda_{u_3}$, $\lambda_{f_1}$ and  $\lambda_{f_3}$ are all positive penalty parameters. The $\widehat{c}_{11}$ and $\widehat{c}_{12}$ can be recovered by optimizing the loss function $J_{\widehat{c}_{11}, \widehat{c}_{12}}(\Theta)$ as discussed before.

\end{itemize}

\begin{figure}[!t]
\centering
\subfloat[]{\includegraphics[trim=0cm 0cm 0cm 1cm,scale=0.375]{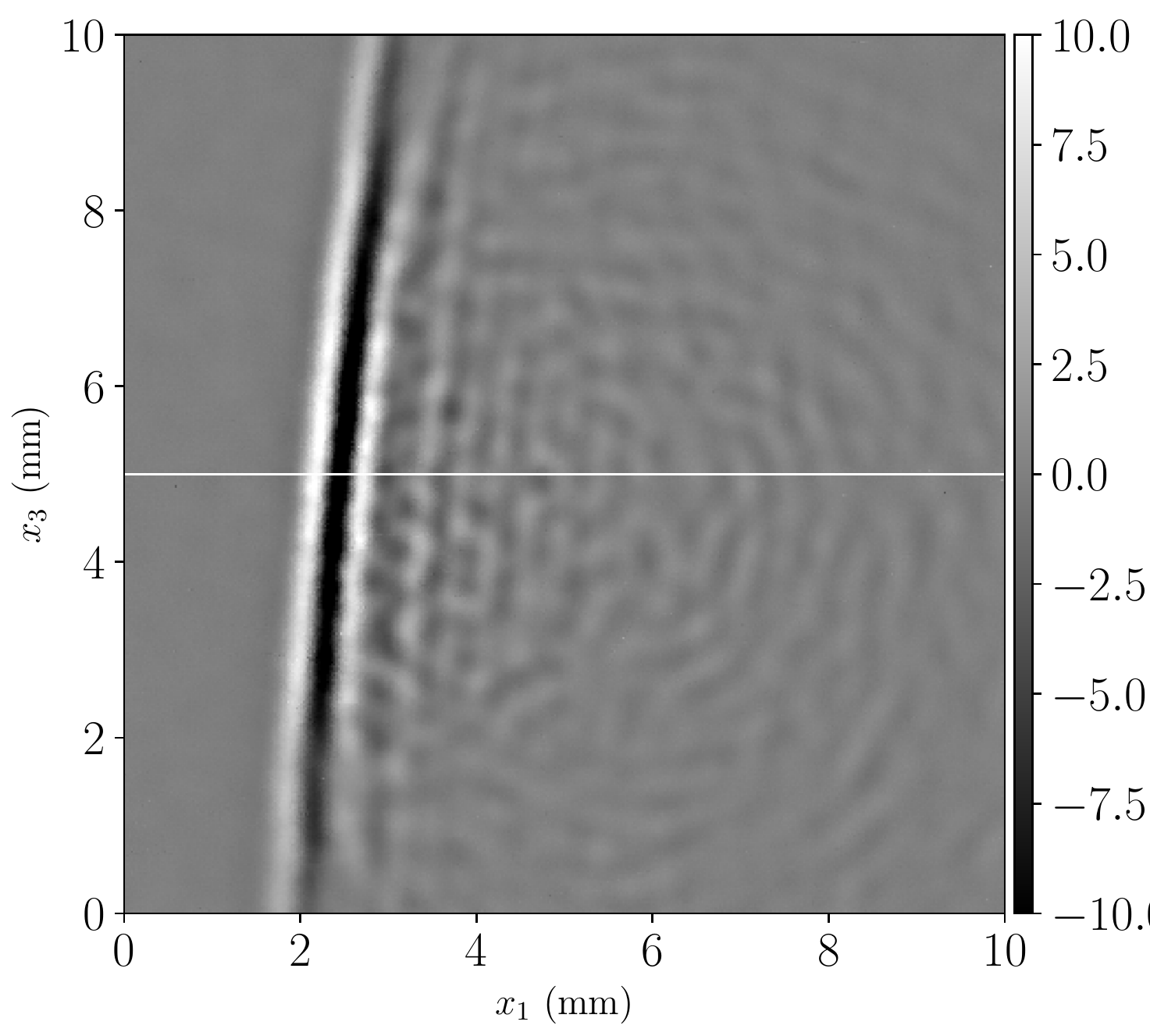}
\label{Figure_3a}}~~~~~
\subfloat[]{\includegraphics[trim=1cm 0cm 2cm 1cm, scale=0.37]{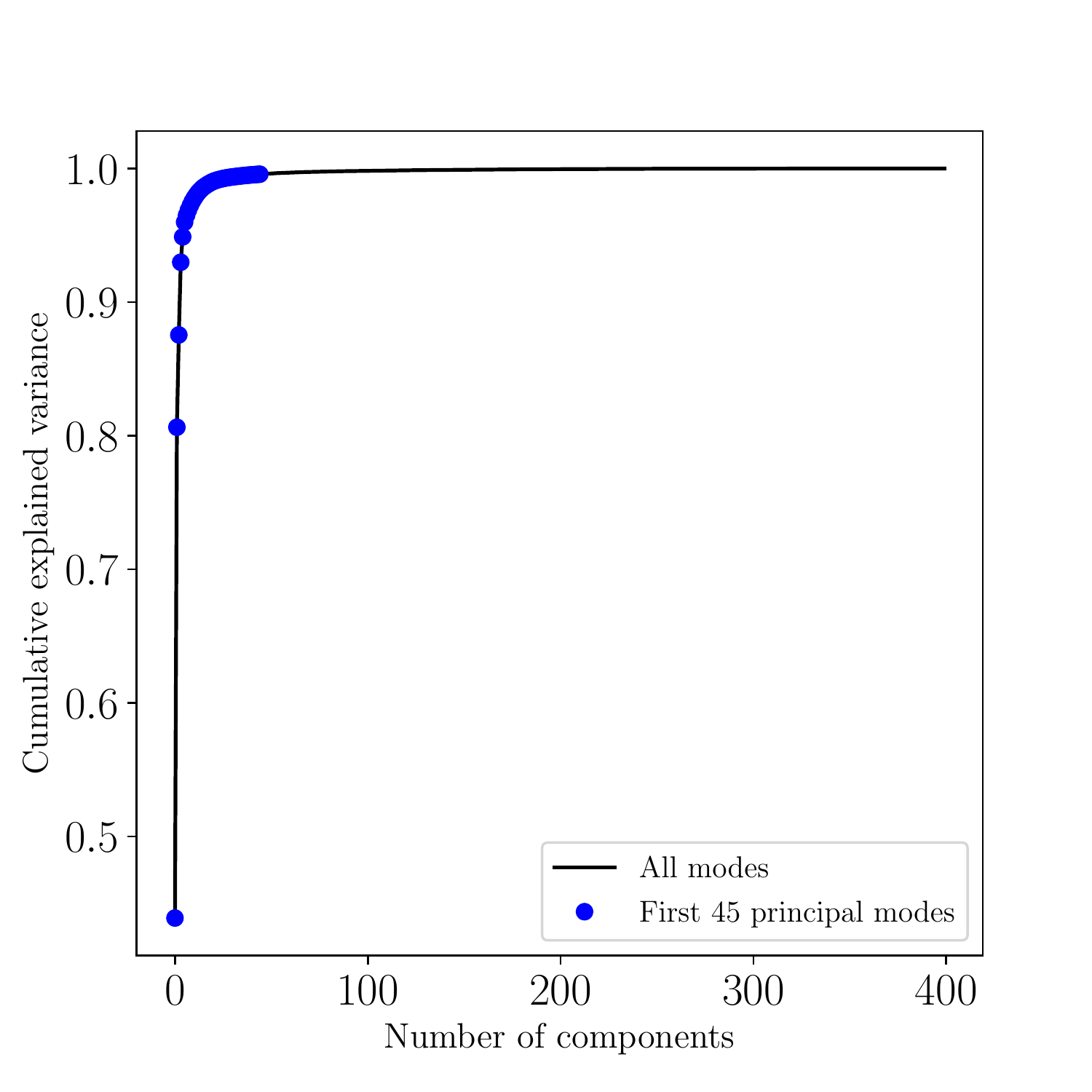}
\label{Figure_3b}}~~~~~~~~~~~
\subfloat[]{\includegraphics[trim=2cm 0cm 0cm 1cm, scale=0.37]{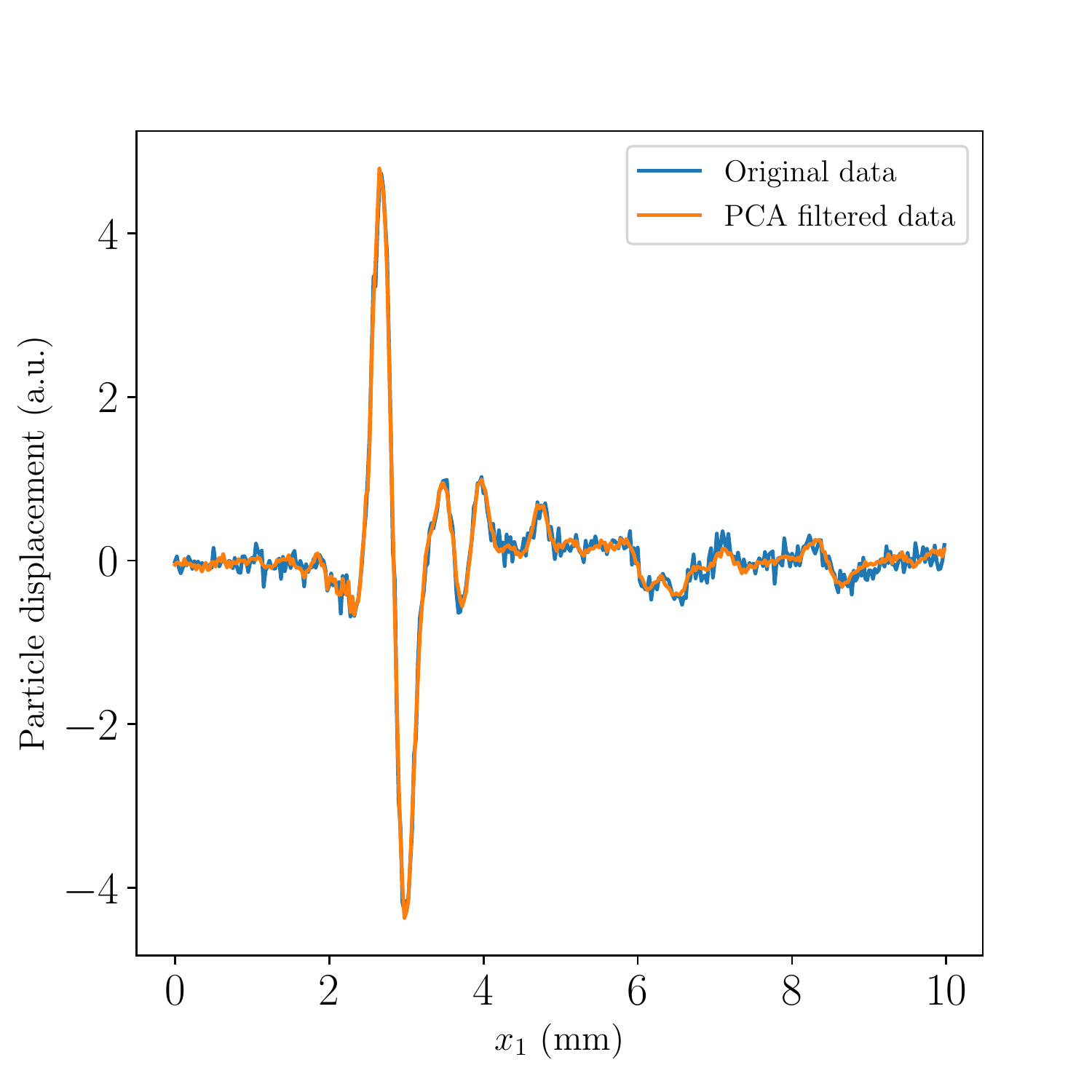}
\label{Figure_3c}}
\label{PCOMP}
\caption{Pre-processing of wavefield data acquired at 5 MHz. (a) represents a time snapshot $(t = 3.84~\mu s)$ of experimental raw wavefield data $u_2$, acquired at frequency of 5 MHz with $N_{x_1}=N_{x_3}=400$.
(b) represents cumulative explained variance of (a) (plotted in black solid line), which also shows the first 45 modes containing the significant energy (plotted in blue solid markers). (c) represents a comparison between raw and filtered trace extracted at $x_3=5~\text{mm}$ marked by the solid white line in Figure 3a.}
\label{precond}
\end{figure}

\section{Pre-processing of experimental data}
In this study, we have analyzed two sets of wavefield data acquired over the Nickel metal surface of dimension $10~\text{mm}\times 10~\text{mm}$ with $(50 \times 50)$ and $(400 \times 400)$ grids. Wavefield data for both grids are based on the collection of 384 time snapshots $(\text{with time step}~dt=0.02~\mu s)$ stored in a 3D array of size $[N_{x_1}, N_{x_3}, N_t]=[50, 50, 384]$ and $[N_{x_1}, N_{x_3}, N_t]=[400, 400, 384]$, respectively.  
To filter out these very small energy events, we employed  the method of \textit{principal component analysis (PCA)},  which is a method for extracting the variance structure from the high-dimensional data and project the data into a subspace such that the variance of projected data is maximized. The PCA based analysis confirms that the wavefield data for $(400 \times 400)$ has some high-frequency and very small energy events, which will result in a slow convergence of PINNs due to spectral bias in the neural network, i.e., the highest frequencies are learned last and hence it takes very long time to resolve them \cite{spbias}.

The first step in the  implementation of PCA is to compute the singular value decomposition.
We represent a time snapshot with $\bm{X}$ and its singular value decomposition as
\begin{align} \label{pcom}
\bm{X} = \bm{U}\bm{S}\bm{V}^\intercal,  \end{align}
where $\bm{U}$ is a unitary matrix and $\bm{S}$ is the diagonal matrix of singular values $s_i$ and $\bm{V}$ is matrix of eigenvector of covariance matrix $\myfrac{\bm{X}^\intercal \bm{X}}{(N_{x_1} -1)}$.  It is to be noted that $\bm{X}$ does not need to be a centered matrix, i.e., symmetric and idempotent.

In Equation (\ref{pcom}), the term $\bm{US}$ represents principal components and $\bm{V}$ represent principal direction.

To filter out the data corresponding to low energy principal components, a ratio of the variance of that principal component and the sum of the variances of all individual principal components is computed. Hence, the filtering process employs the zeroing-out of one or more of the smallest principle components by preserving the maximum data variance. Figure \ref{Figure_3a} shows a time snapshot of raw wavefield data $(u_2)$ at a representative time $t=3.84 ~\mu \text{s}$. Figure 3b shows the ratio between the variance of that principal component  and the sum of variances of all individual principal components for the wavefield shown in Figure 3a. Figure 3b shows that the dominant modes of energy are  contained only in the first 45 principal components (blue solid markers), and therefore we filtered out the remaining components. A comparison between filtered and non-filtered traces is shown in Figure \ref{Figure_3c}. The trace is extracted along the solid white horizontal line as shown in Figure 3a. Figure 3b clearly shows that PCA is very effective in filtering out unwanted high-frequency and low energy noises. The choice of first 45 components is very conservative to avoid the loss of any important information. The PCA has been applied to each time snapshot individually on the dataset acquired for $400 \times 400$ grid points. However, the PCA has not been applied on $50 \times 50$ datasets due to the uniform distribution of energy over all the principal components of the time snapshots.

\begin{figure}[!t]
\centering
\subfloat[]{\includegraphics[trim=2cm 0cm 0cm 0cm, scale=0.63]{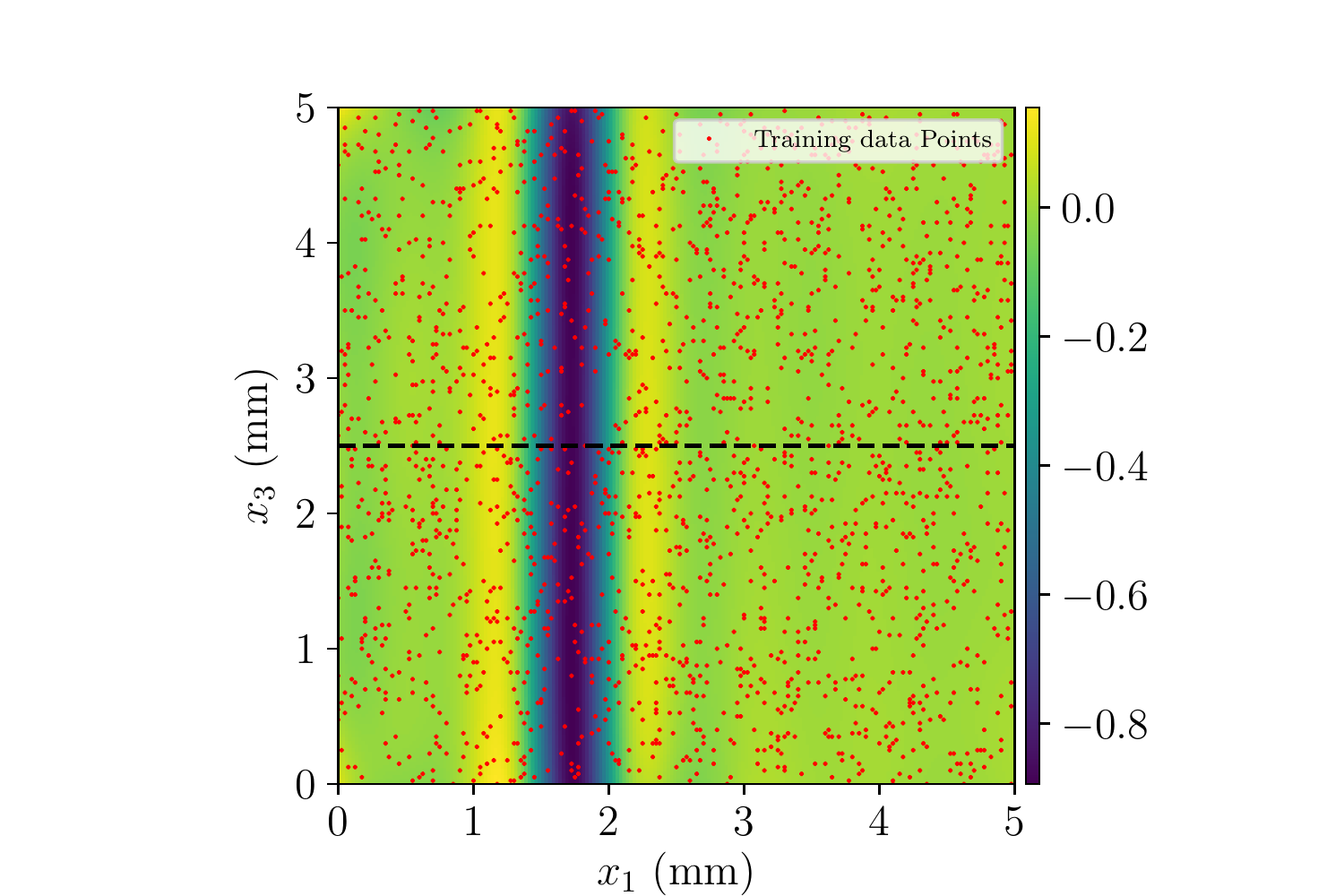}
\label{Figure_3a.pdf}}~~~~
\subfloat[] {\includegraphics[trim=2cm 0cm 0cm 2cm, scale=0.63]{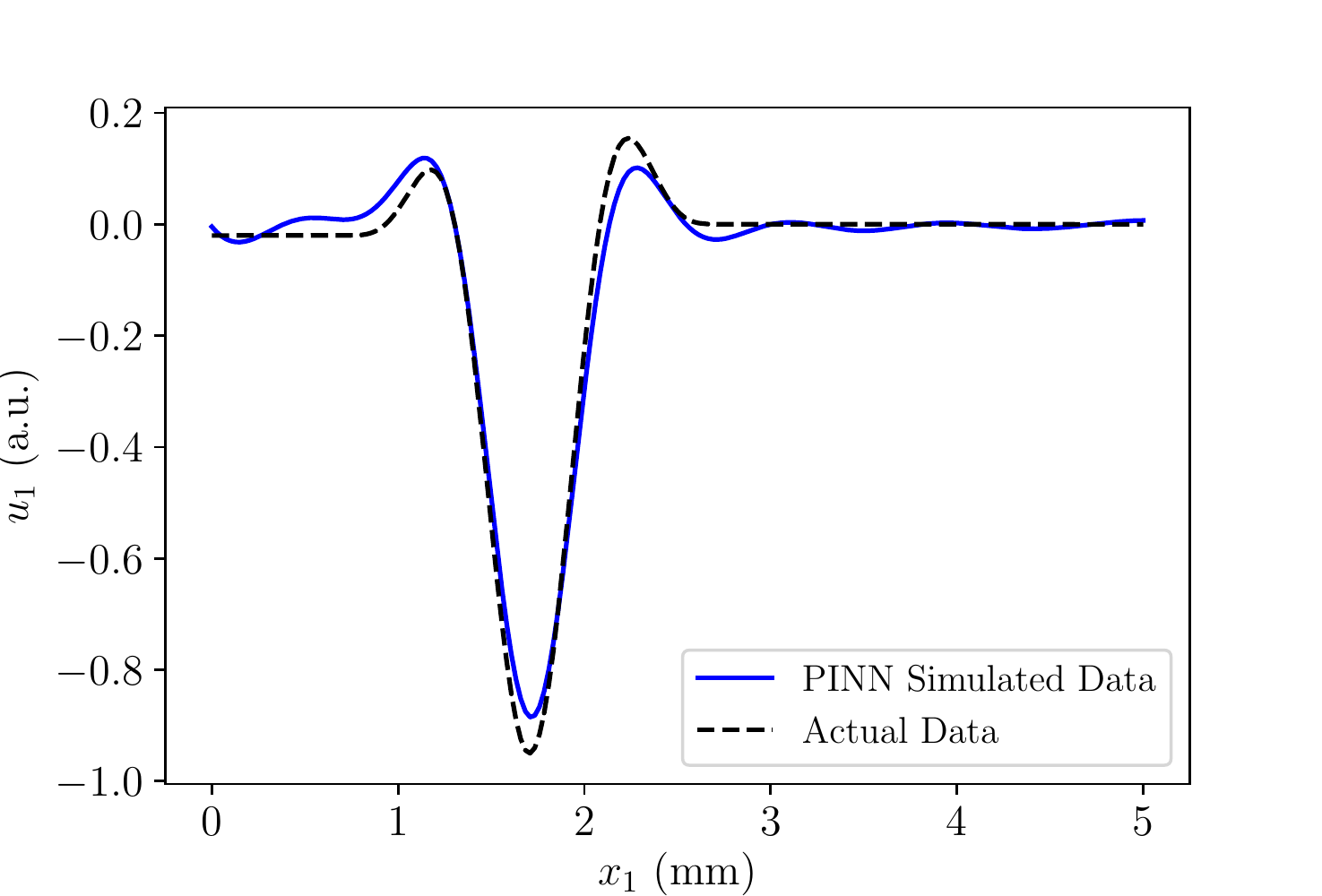}
\label{Figure_3b.pdf}}\\
\subfloat[]{\includegraphics[trim=1cm 0cm 0.5cm 0cm,scale=0.6]{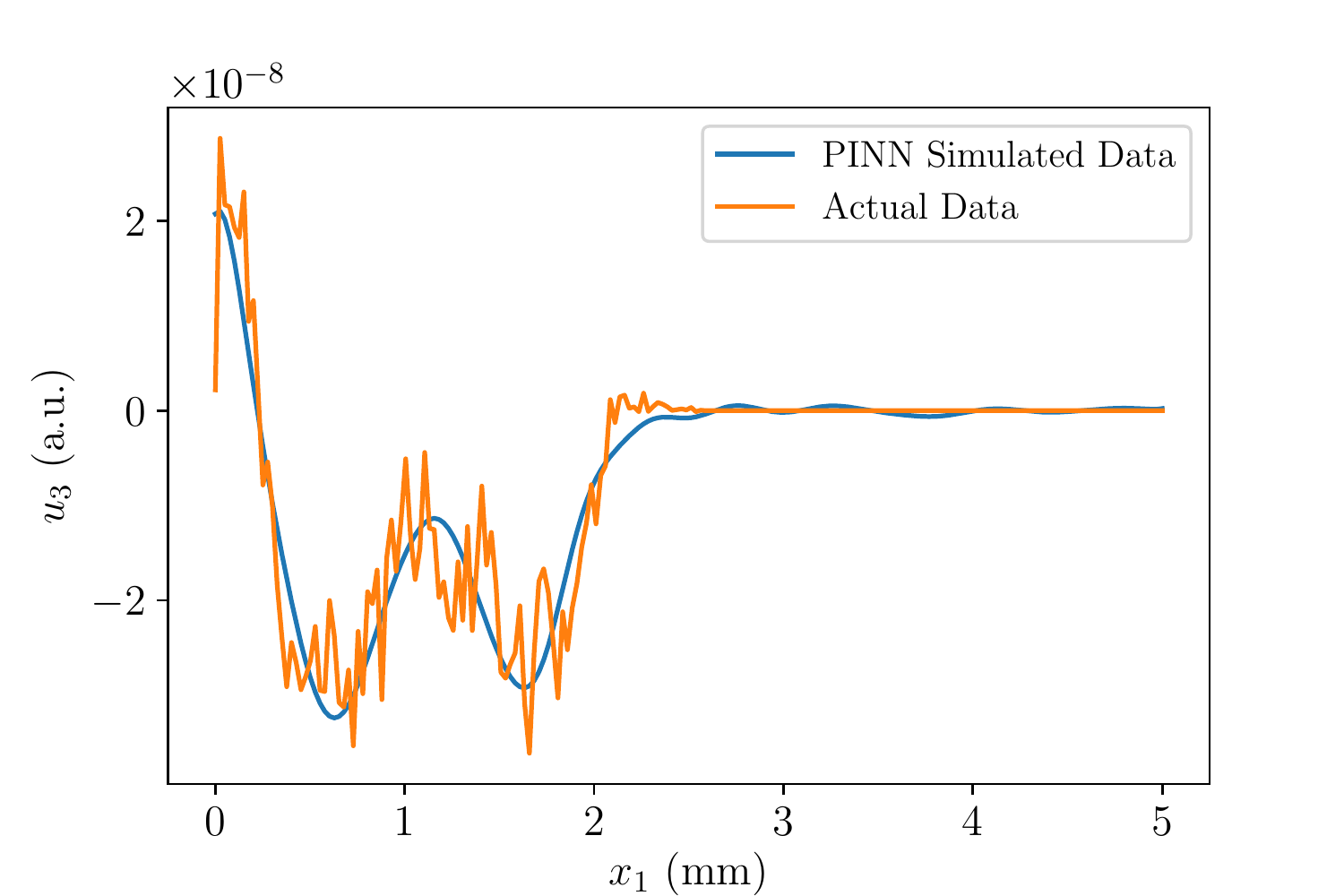}
\label{Figure_3c.pdf}}
\subfloat[]{\includegraphics[trim=1cm 0cm 0.5cm 0cm,scale=0.6]{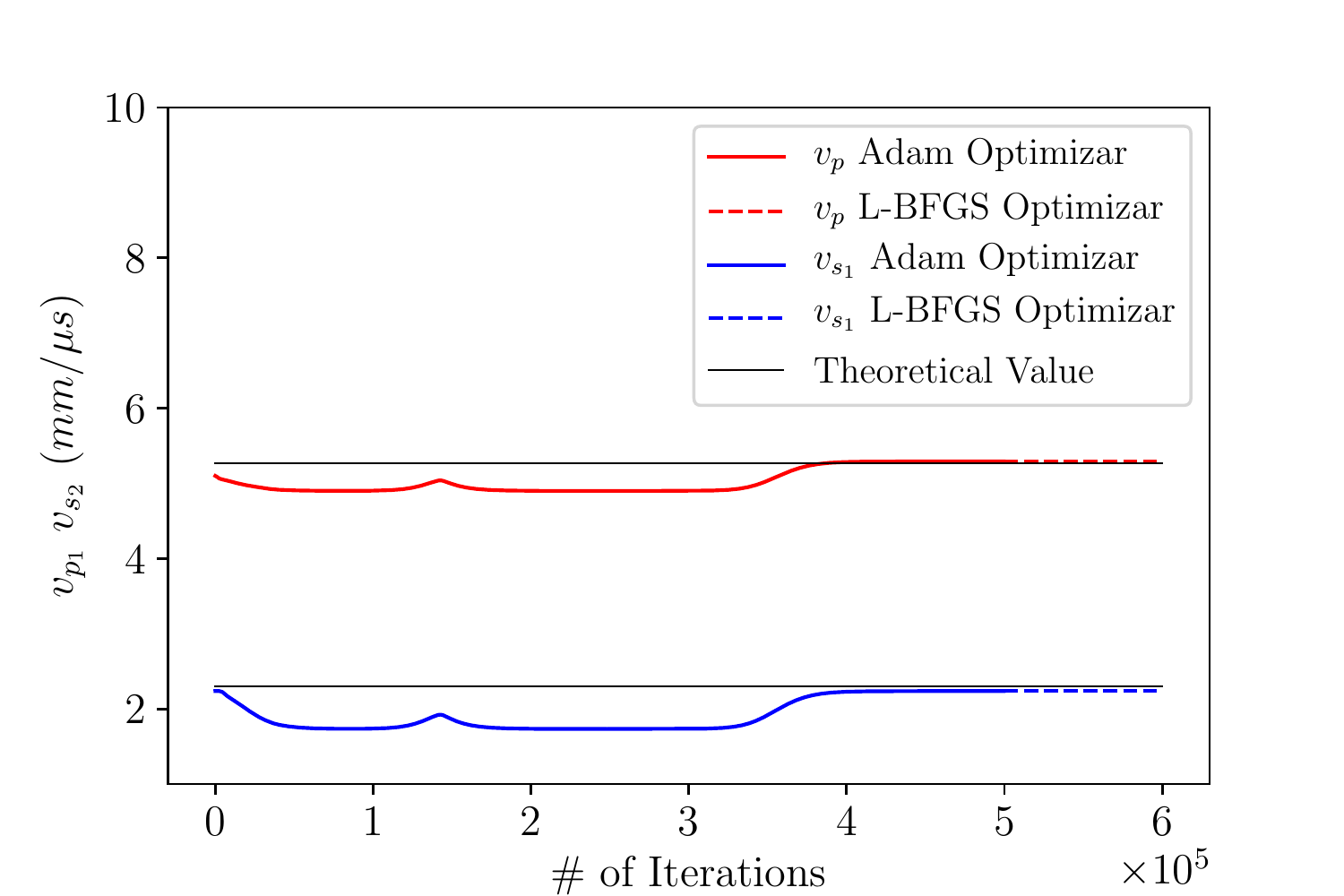}
\label{Figure_3d.pdf}}\\
\caption{Results from training of PINNs for data generated from finite element modeling approach by solving the elastic wave equation for $u_1$ and $u_3$. The data is simulated for a single-crystal Nickel. (a) shows a time snapshot of $u_1$  at $t=0.52~\mu s$. Red solid dots represent the location of data points sampled for the training of PINNs. (b) shows a comparison of actual and PINN simulated $u_1$ along a black dashed horizontal line in (a). (c) shows a comparison of actual and PINN simulated $u_3$ along a black dashed horizontal line in (a), and (d) shows the velocity of longitudinal $(v_{p_1})$ and the first shear wave $(v_{s_1})$ along with the actual velocities plotted in solid black line.}  \label{Figure_4}
\end{figure}

\begin{table}
\centering

	\label{Table1}
	\caption {${c_{11}}$ and ${c_{12}}$ of single-crystal Nickel}
	\begin{tabular}{|c|c|c|}
		\hline
		\textbf{} & $\bm{c}_{11}$ (GPa) & $\bm{c}_{12}$ (GPa)\\
		\hline
		Actual values & 257  &  153\\
		\hline
		PINN computed values & 249.20  & 160.20 \\
		\hline
		Relative $L_2$ errors & 3\%  & 4\% \\
		\hline
	\end{tabular}
\end{table}

\begin{table}
 \centering

	\label{Table2}
	\caption{$v_{s_2}$ and $c_{44}$ of polycrystalline Nickel with variations in the width and depth of the network.}
	
	\begin{tabular}{|c|c|c|}
		\hline
		\textbf{Width (N), Depth (L)} & $v_{s_2}~\text{mm} \mu s~(\max, \min)$ & $c_{44}~$ GPa $(\max, \min)$ \\
		\hline
		N=50,~~ L=2 & $(2.98, 2.99)$ & (79.04, 79.57) \\
		N=50,~~ L=3 & $(2.98, 2.99)$ & (79.04, 79.57)  \\
		N=50,~~ L=4 & $(2.96, 2.99)$ & (77.98, 79.57)  \\
		N=50,~~ L=5 & $(2.94, 3.00)$ & (76.93, 80.10) \\
		N=50,~~ L=6 & $(2.95, 3.00)$ & (77.45, 80.10) \\
		\hline
		N=10,~~ L=6 & $(2.98, 2.99)$ & (79.04, 79.57)  \\
		N=20,~~ L=6 & $(2.98, 3.09)$ & (79.04, 80.63)  \\
		N=30,~~ L=6 & $(2.94, 3.00)$ & (76.93, 80.10)\\
	    N=40,~~ L=6 & $(2.94, 3.00)$ & (76.93, 80.10)\\
		N=50,~~ L=6 & $(2.95, 3.00)$ & (77.45, 80.10) \\
		\hline
	\end{tabular}
\end{table}

\begin{figure}[!t]
\centering
\subfloat[]{\includegraphics[width=0.43\textwidth]{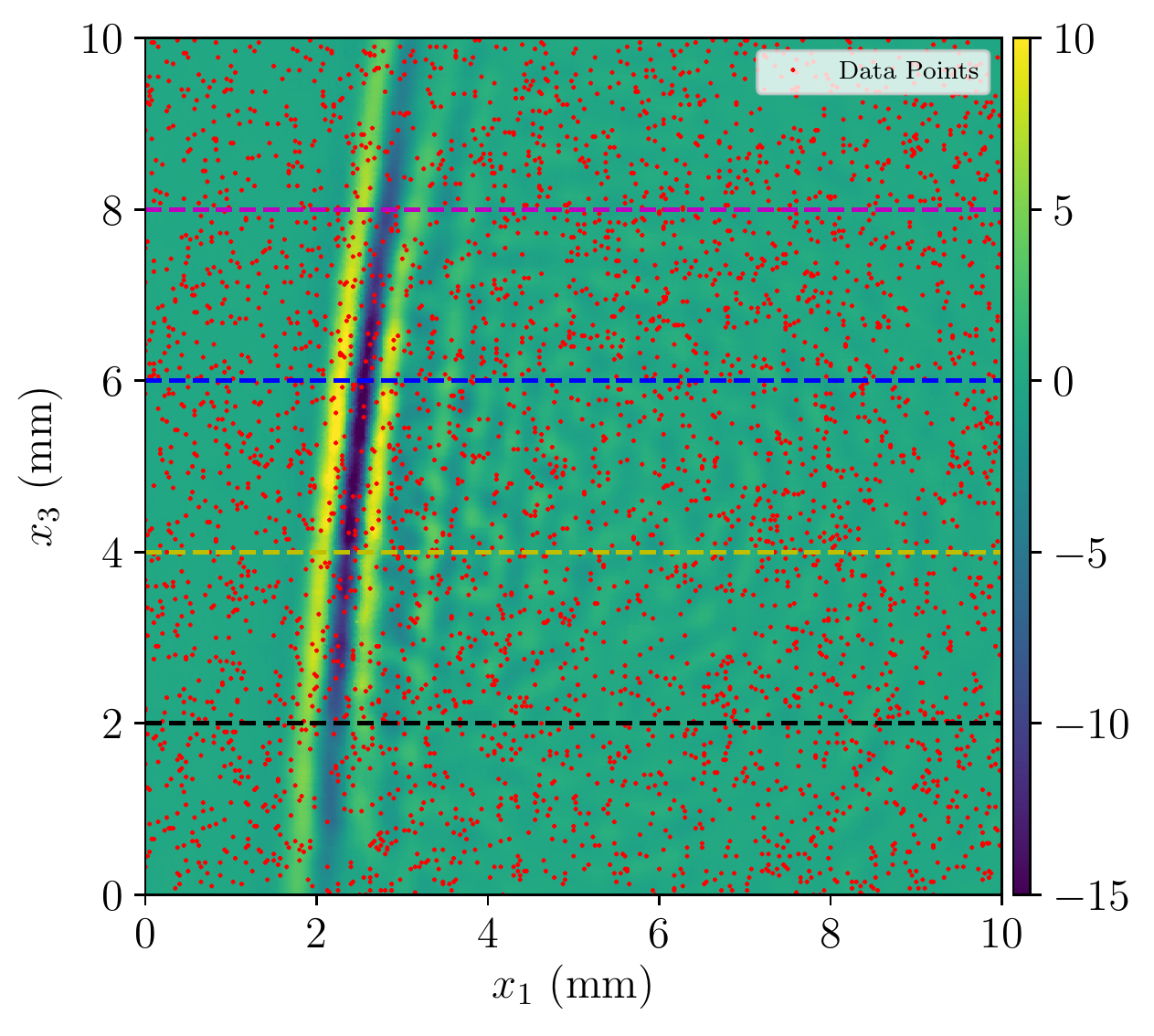}
\label{Figure_4a}}
\subfloat[]{\includegraphics[width=0.385\textwidth]{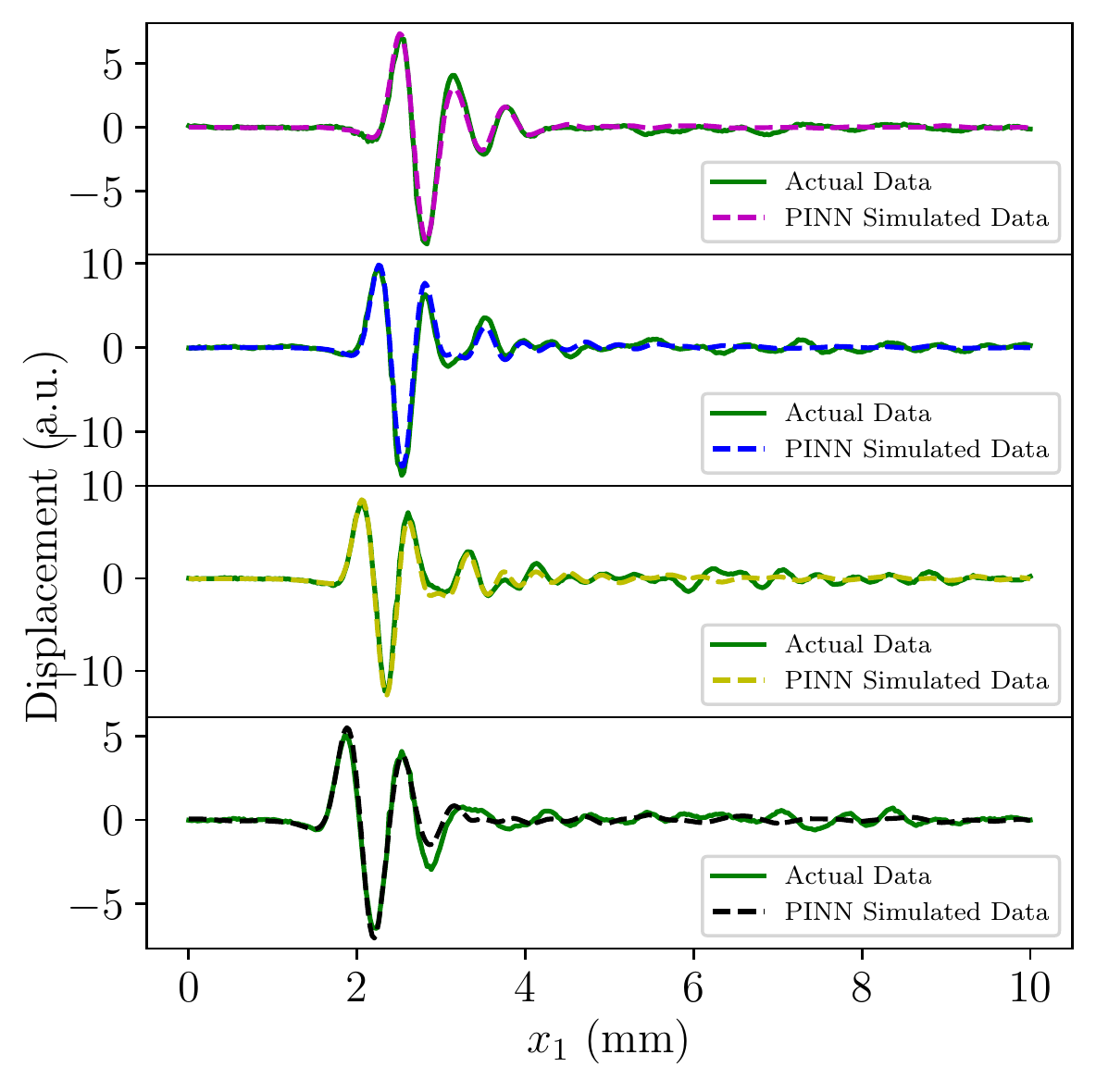}
\label{Figure_4b.png}}\\
\subfloat[]{\includegraphics[width=0.43\textwidth]{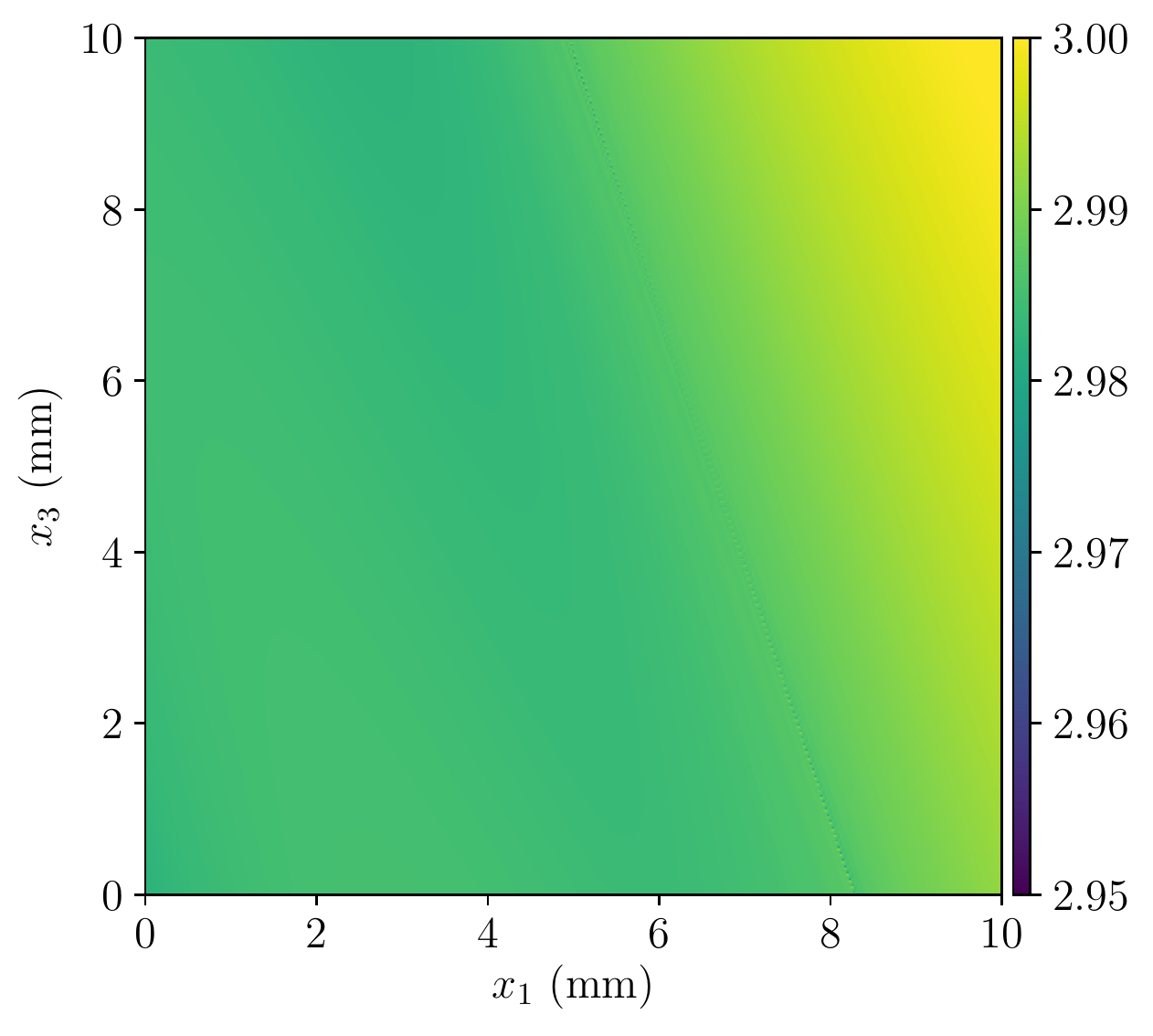}
\label{Figure_4c}}
\subfloat[]{\includegraphics[width=0.43\textwidth]{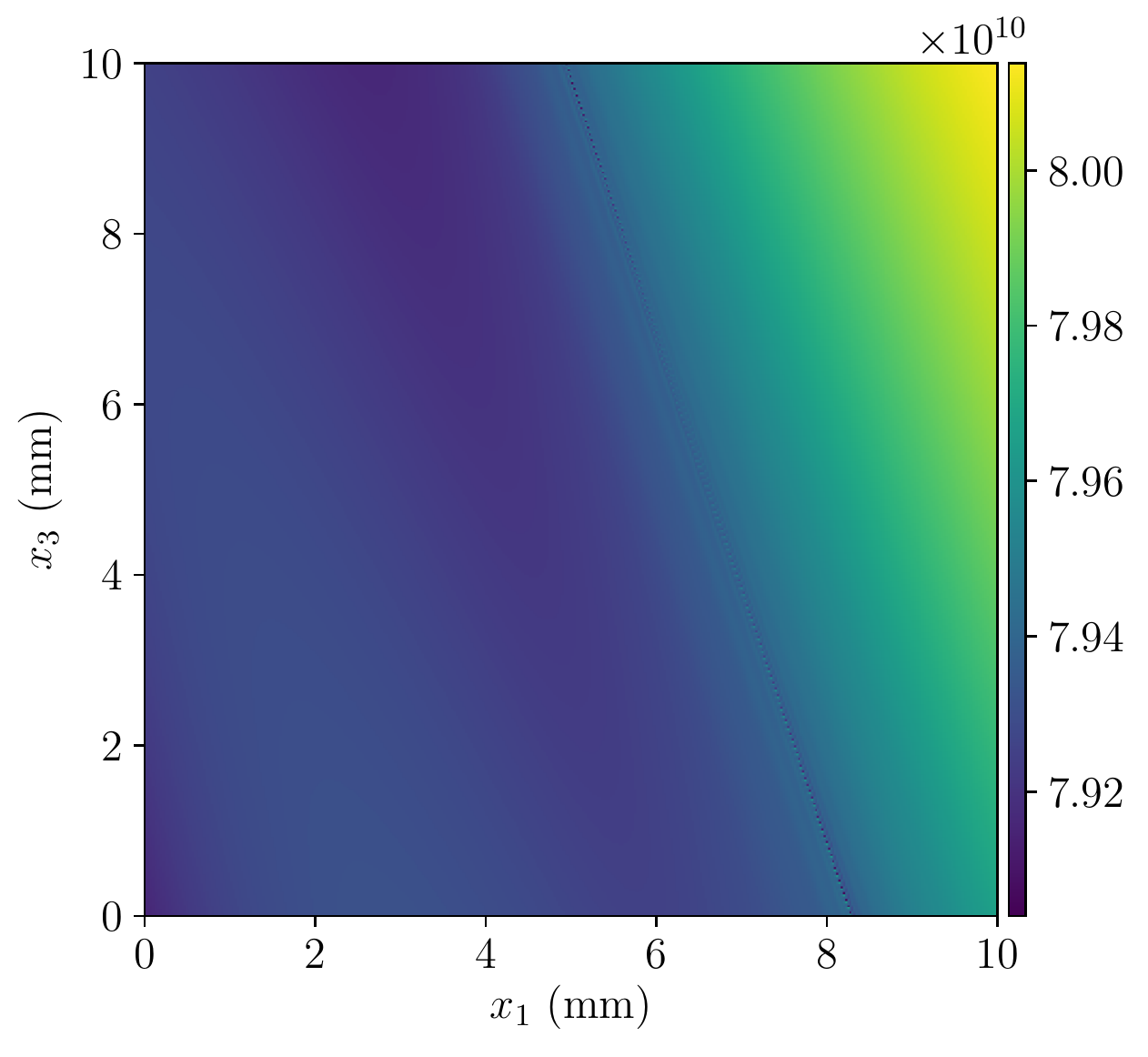}
\label{Figure_4d.png}}
\caption{Results from training of PINN for wavefield data $(u_2)$ acquired over the surface of a metal plate of dimension $(10~\text{mm} \times 10~\text{mm})$ with $(N_{x_1},~N_{x_3})=(400 \times 400)$ spatial grid points. (a) shows a time snapshot of PCA filter data at $t=3.84~\mu s$ overlaid with red colored dots representing the location of training data points for PINNs. (b) shows a comparison between the traces of actual and PINN simulated data along the color lines shown in (a). Color of traces corresponds to the color and the location of lines in (a). (c) shows the speed of space-dependent second shear wave mode $v_{s_2}(x_1,x_3)$. Spatial variation of speed confirms the distribution of polycrystals in the material. (d) shows the spatial variation of $c_{44}$.}  
\label{c44_400}
\end{figure}

\begin{table}
 \centering
	\label{Table3}
	\caption{Hyper-parameters of the neural network and the training data details}
	
	\begin{tabular}{|c|c|c|c|c|c|c|}
		\hline
		\multicolumn{7}{|c|}{\textbf{Hyperparameters used in this study and the details of training data}} \\
		\hline
		\textbf{Results in} & \textbf{Depth} & \textbf{Width} & \textbf{Number of iterations} & \textbf{Activation} & \textbf{Learning} & \textbf{\% of spatial } \\
		 &  &  &  $\mathbf{( \times 10^5 )}$ & \textbf{function} & \textbf{rate} & \textbf{data used} \\
		\hline
		Figure 4   &  2 & 32  & 2 & $\sin$ & $1\text{e}-4$ & 40\% \\
		\hline
		Figure 5 NN1   &  6 & 50  & 2 & $\sin$ & $2\text{e}-4$ & 20\% \\
		\hline
		Figure 5 NN2 $(\widehat{c}_{44})$   &  2 & 16  & 2 & $\sin$ & $2\text{e}-4$ & 20\% \\
		\hline
	\end{tabular}
\end{table}

\section{Results}
In this section, we will discuss the efficacy of PINN, described in Section \ref{PINN}, in quantifying the microstructure of polycrystalline Nickel. We primarily considered the two cases, first, we validated the results of PINN by computing the $\widehat{c}_{11}$ and $\widehat{c}_{12}$ for a single-crystal Nickel. In the second case, the spatial variation of $\widehat{c}_{44}$ is computed for the polycrystalline Nickel using PINN.

\subsection{PINN for single-crystal Nickel}
To validate the applicability of PINN for microstructure quantification, computation of $\widehat{c}_{11}$ and $\widehat{c}_{12}$ is performed for an isotropic single-crystal Nickel of dimension $5~\text{mm} \times 5~\text{mm}$ with crystallographic orientation of $(0, 0, 0)$. The crystallographic orientation describes the angle between the plane described by the Cartesian system and plane-of-symmetry of the material. The stiffness tensor $\bm{C}$ for the numerical simulation is adopted from Xu \etal \cite{xu}. The data  for $u_1$  and $u_3$ were generated numerically by solving the elastic wave equation \cite{noro} for a plane wave of $5~\text{MHz}$ as an initial condition. The spatial and temporal grid size for discretization are $[\Delta x_1,~\Delta x_3,~\Delta t] = [0.025~\text{mm}, 0.025~\text{mm}, 0.40~\text{ns}]$. For training of the PINN, we considered 14 time snapshots representing the time duration between $0.4-0.57~\mu s$, and each time snapshot is spatially sampled with $40 \%$ data. Figure 4a represents one of the time snapshots at $t=0.52~\mu s$ overlaid with the red solid dots representing the training data points sampled in an \textit{i.i.d} (independent and identical distribution) way. Thereafter, PINN simulation is performed, as described in Section \ref{PINN}, to recover the full field of $u_1$ and $u_3$, $\widehat{c}_{11}$ and $\widehat{c}_{12}$. The hyper-parameters of the neural networks are given in Table III. 

The values of $\lambda_{u_1},~\lambda_{u_2},~\lambda_{f_1},~\text{and}~\lambda_{f_3}$ in (\ref{eq8}) are taken as 50, 20, 1, and 1, respectively. Figures 4b and 4c show a comparison between the actual and PINN simulated solutions of $u_1$ and $u_3$, extracted along the black dashed line shown in Figure 4a. The comparison of $u_1$ shows a very good agreement. Amplitudes of $u_3$ are very small $(O(10^{-8}))$ and are represented by a very high frequency spatial variation as shown in Figure 4c. The small amplitudes of $u_3$ are defined by the wave vector defining the direction of propagation of wave energy, which is directed towards $x_1$ for this case. Therefore, a very small fraction of energy is responsible for causing the oscillation in $x_3$ direction, resulting in the small scales of $u_3$. However, pairing PINNs with appropriate scaling factor results into a reasonable approximation of $u_3$ as shown in Figure 4c. Figure 4d represents $v_{p_1}$ and $v_{s_1}$ recovered from PINN. In Figure 4d, the solid and dashed portions of red and blue lines correspond to Adam and L-BFGS optimizer, respectively. We present the speed of waves for in-plane particle motion computed from (\ref{speed}) along with the actual values. The theoretical and PINN inferred values of $v_{p_1}$ are $5.27~{\text{mm}}/{\mu s}$ \cite{xu} and $5.29~{\text{mm}}/{\mu s}$, respectively. This gives a relative error of $0.3\%$. However a relative error of $2.6\%$ is observed for $v_{s_1}$ with actual and PINN inferred value being $2.30~{\text{mm}}/{\mu s}$ and $2.23~{\text{mm}}/{\mu s}$, respectively. 
We also present the values of $c_{11}$ and $c_{12}$ in Table I, which are computed from PINN. Table I also shows the actual values $c_{11}$ and $c_{12}$ and $L^2-$ relative error between actual and PINN computed value for single-crystal Nickel. The relative errors are very small and amount to 3\% and 4\% for $c_{11}$ and $c_{12}$, respectively.

\subsection{PINN for polycrystalline Nickel}
Next, we show the capabilities of PINN in inferring $\widehat{c}_{44}$ with the experimental wavefield data measured as $u_2$. The computation of $\widehat{c}_{44}$ is carried out for wavefield data acquired with $[400 \times 400]$ and $[50 \times 50]$ grids in space. A total of 120 time snapshots, representing wavefield data from $2~\mu s$ to $4.42~\mu s$, are selected. Thereafter, $12.5 \%$ of time snapshots are selected in the temporal domain and $20\%$ data points were sampled spatially in each time snapshot using an \textit{i.i.d.} approach. Choosing the optimal hyper-parameters is important for the convergence of neural network training. A more sophisticated meta-learning approach can be used to obtain the best hyper-parameters of the networks, but due to high computational cost associated with it, we choose the network architecture based on the complexity of the data and our past experience with PINNs. Nevertheless, Table II  gives the maximum and minimum values of $c_{44}$ and $v_{s_2}$ for the polycrystalline Nickel obtained using different width and depth of the networks. Table II clearly shows that parameters become consistent for depth and width of 6 and 50, respectively. Therefore, the hyper-parameters used in this study can be considered as adequate if not optimal.

Figure 5a shows a time snapshot of $u_2$ at $t=3.84~\mu s$ and overlaid with red solid dot representing the samples chosen for training of PINN. Subsequently, PINN simulations are performed with hyper-parameters shown in Table III. Figure 5b shows the comparison between actual and PINN simulated data along for $x_3=2~\text{mm}, 4~\text{mm}, 6~\text{mm},$ and $8~\text{mm}$ represented by lines in black, yellow, blue and magenta colors in Figure 5a, respectively. Figure 5b shows a very good agreement between actual and PINN simulated solutions. Figure 5c shows the spatial variation of $v_{s_2}$ in a range of $[2.95~{\text{mm}}/{\mu s}, 3.00~{\text{mm}}/{\mu s}]$ ensuring the change in stiffness of crystals properties, especially rigidity $(c_{44})$. We note that for a homogeneous single-crystal Nickel, the speed of $v_{s_2}$ is $2.97~{\text{mm}}/{\mu s}$ \cite{DUS}.  Figure 5d shows the variation $c_{44}$ of the polycrystalline Nickel, which varies from $79~\text{GPa}$ to $82~\text{GPa}$ and conforms with the $v_{s_2}$ as low and high speeds correspond to low and high value $c_{44}$.     

Next, we compute the uncertainty in the $v_{s_2} (\widehat{c}_{44})$ by using the wavefield data for $u_2$ acquired with $[50 \times 50]$ grid points in space. To perform the PINN simulation we utilized unfiltered PCA data comprising 120 time snapshots from  $2~\mu s$ to $4.42~\mu s$ and sampled with a rate of $12.5\%$ in the temporal domain and $60\%$ in the spatial domain. The slow convergence of PINN due to presence of noise in the data is circumvented by the approach of weight decay \cite{wd}, which eventually bounds the weights of neural network, hence resulting in a faster convergence. Therefore, the loss function in equation (\ref{eq5}) is further modified and expressed as
\begin{align} \label{eq5mod}
J_{\widehat{c}_{44}}(\Theta)= \lambda_{u_2} MSE_{u_2} + \lambda_{f_2} MSE_{f_2} +\underbrace{\lambda_{\bm{w}} ||\bm{w}||_2^2}_{\text{Weight decay}}.
\end{align}

The PINN for $u_2$ is executed five times with different initialization of parameters of neural network and optimized loss function as defined in $(\ref{eq5mod}$). The values of $\lambda_{u_2},~\lambda_{f_2},~\text{and}~\lambda_{\bm{w}}$ are taken as $50,~5,~\text{and}~0.1$, respectively. The standard deviation and mean of minimum and maximum value of $v_{s_2}$, inferred from the PINNs, are $(3 \times 10^{-3}~\text{mm}/\mu s,~4 \times 10^{-4}~\text{mm}/\mu s)$ and $(2.9831~\text{mm}/\mu s,~ 2.9845~\text{mm}/\mu s)$ . The small standard deviation confirms the consistency of the simulations carried out by the PINNs. Additionally, due to coarse sampling, the speed of $v_{s_2}$ observed from  this data set does not show large spatial variation as it is not able to capture the fine variation of grains.

\section{Summary}
In this study, we quantified the  microstructure as defined by elements of stiffness tensor $(\widehat{c}_{ij})$, of a cubically symmetric polycrystalline Nickel. The $(\widehat{c}_{ij})$  are inferred by using the physics-informed neural network (PINN) trained with the ultrasonic wavefield dataset acquired at the frequency of 5 MHz. The PINNs are informed with two sets of decoupled second order hyperbolic partial differential equations, each defining the in-plane and out-of-plane particle displacement, respectively. Data of particle displacements are recorded in a spatio-temporal way and represented by three orthogonal components of displacement; $u_1$ and $u_3$ for in-plane and $u_2$ for out-of-plane motion. We validate the efficacy of PINNs by first computing the $\widehat{c}_{ij}$ for a single-crystal and polycrystal Nickels. For a single-crystal Nickel, a shallow network of two hidden-layers produced very accurate estimations of $\widehat{c}_{11}$ and $\widehat{c}_{12}$ (represented by speeds of longitudinal $(v_{p_1})$ and first shear waves $(v_{s_1})$, respectively). Subsequently, PINNs are implemented for polycrystal Nickel and trained with experimental datasets representing $u_2$. We used two experimental wavefield imaging datasets acquired at different spatial resolutions of $400\times400$ and $50\times 50$ grid points. A comparison between actual and PINNs simulated data shows a very good agreement and spatial variation  of $\widehat{c}_{44}$ (representing $v_{s_2}$, speed of second shear wave) shows presence of polycrystal of different stiffness. We used the second dataset acquired at $50\times 50$ grid points to compute the uncertainly and consistency of PINNs. A very small standard deviation in minimum and maximum $v_{s_2}$ proves the consistency of PINNs. All the above computational experiments have used a very small fraction of data in space $(40\% ~\text{and}~ 20\%)$ and time $(12.5\%)$.

In this work we have used an \textit{i.i.d.} approach for sampling the data points. However, there are other approaches like importance sampling  that can be used with sparse dataset.
In this study, a generic representation of PINNs works very accurately but we observed a slow convergence for some high frequencies, which is generally caused by \textit{spectral bias} \cite{sifan} phenomena present in the neural network. Therefore, the future scope of work will include embedding the input variables $(\bm{x})$ in Fourier space and performing the training through Fourier feature mappings corresponding to individual frequencies \cite{smu}. Additionally, we presented the quantification of microstructure based on the 2D wavefield datasets for a single ultrasonic frequency. Future ongoing work will include extension of the current work for 3D datasets acquired for various frequencies.      

\section*{Acknowledgments}
The work was supported by OSD/AFOSR MURI Grant FA9550-20-1-0358, DOE PhILMs project (DE-SC0019453) and the DURIP Grant W911-NF-2110089. This research was conducted using computational resources and services at the Center for Computation and Visualization (OSCAR), Brown University, Oak Ridge National Lab (ORNL) and Longhorn machine of Texas Advanced Computing Center's (TACC). KS would like to acknowledge Dr. Helen Kershaw of NCAR, Colorado, for providing help with problems concerning the computation.

\vspace*{-3cm}
\begin{IEEEbiographynophoto}{Khemraj Shukla}
is an Assistant Professor (Research) in the Division of Applied Mathematics at Brown University, USA. His research focuses on the development of scalable codes on heterogeneous computing architectures. 
\end{IEEEbiographynophoto}
\vspace*{-3cm}
\begin{IEEEbiographynophoto}{Ameya D. Jagtap}
is an Assistant Professor of Applied Mathematics (Research) in the Division of Applied Mathematics at Brown University, USA. His research focuses on the intersection of machine learning and computational physics. 
\end{IEEEbiographynophoto}
\vspace*{-3cm}
\begin{IEEEbiographynophoto}{James L Blackshire}
is a Senior Research Engineer and Senior Material Scientist at the Wright-Patterson Air Force Research Laboratory, based out of Ohio, USA. 
\end{IEEEbiographynophoto}
\vspace*{-3cm}
\begin{IEEEbiographynophoto}{Daniel Sparkman}
is a Scientist at Wright-Patterson Air Force Research Laboratory, based out of Ohio, USA. 
\end{IEEEbiographynophoto}
\vspace*{-3cm}
\begin{IEEEbiographynophoto}{George Em Karniadkais}
is the Charles Pitts Robinson and John Palmer Barstow Professor of Applied Mathematics and Engineering at Brown University, USA, and he has a joint appointment with PNNL, where he is the Director of the PhILMs Collaboratory on physics-informed learning.
\end{IEEEbiographynophoto}




\end{document}